\documentclass[12pt, a4paper]{article}
\usepackage{graphicx}
\usepackage{amssymb}
\usepackage{amsmath}
\usepackage{bm}
\usepackage{color}
\usepackage{theorem}

\definecolor{Orange}{cmyk}{0,0.61,0.87,0}
\definecolor{JungleGreen}{cmyk}{0.99,0,0.52,0}
\definecolor{OliveGreen}{cmyk}{0.64,0,0.95,0.40}
\definecolor{Brown}{cmyk}{0,0.81,1,0.60}
\definecolor{RoyalBlue}{cmyk}{0.71,0.53,0,0.12}

\usepackage[colorlinks=true, linkcolor=black, citecolor=black,
urlcolor=black]{hyperref}

\begin{document}

\begin{titlepage}

\begin{flushright}
FTPI-MINN-15/02 \\
UMN--TH--3417/15 \\
IPMU15-0011
\end{flushright}

\vskip 1.35cm
\begin{center}

{\large
{\bf
SU(5) Grand Unification in Pure Gravity Mediation
}
}

\vskip 1.2cm

Jason L. Evans$^a$,
Natsumi Nagata$^{a,b}$,
and
Keith A. Olive$^a$

\vskip 0.4cm

{\it $^a$William I. Fine Theoretical Physics Institute, School of
 Physics and Astronomy, \\ University of Minnesota, Minneapolis, MN 55455,
 USA}\\ [3pt]

{\it $^b$Kavli IPMU (WPI), TODIAS, University of Tokyo, Kashiwa
 277-8583, Japan}

\date{\today}

\vskip 1.5cm

\begin{abstract}

 We discuss the proton lifetime in pure gravity mediation models with
 non-universal Higgs soft masses.  Pure gravity mediation offers a
 simple framework for studying SU(5) grand unified theories with a split
 supersymmetry like spectra.  We find that for much of the parameter
 space gauge coupling unification is quite good leading to rather long
 lifetimes for the proton.  However, for $m_{3/2}\sim 60$~TeV and
 $\tan\beta\sim 4$, for which gauge coupling unification is also good,
 the proton lifetime is short enough that it could be  in reach of
 future experiments.

\end{abstract}

\end{center}
\end{titlepage}

\section{Introduction}

After the initial run of the LHC, the constraints on new physics are
rather severe \cite{lhc}.  Although models can still be made to realize
weak-scale mass spectra, sfermion masses of generic models like the
constrained minimal supersymmetric standard model (CMSSM) \cite{cmssm}
are now required to be larger than about a TeV. As a result, the
naturalness of supersymmetry (SUSY) has been called into question. However, it
was perhaps naive to expect nature to fall into our strict definition of
naturalness with less than $10\%$ fine-tuning. Supersymmetry, with
sfermion masses larger than a TeV, still solves the larger hierarchy
problem associated with grand unification and/or the Plank
scale. Furthermore, if the sfermion masses are set by the gravitino
mass, $m_{3/2}$, and are larger than about 10 TeV, the gravitino
lifetime is short enough that it decays before BBN \cite{gravitinoBBN}. Moreover, as the
mass scale of the sfermions is pushed beyond the weak scale, the
constraints on SUSY models from flavor and CP violation in the sfermion
sector are greatly relaxed \cite{flavor}. These advantages, plus the fact that
sfermion masses this large are consistent with a larger Higgs mass like
the $126$~GeV Higgs boson seen at the LHC \cite{lhch} suggest we relax our strict
definition of naturalness.

Large sfermion masses like those found in split supersymmetry \cite{split} are realized in models such as pure gravity
mediation (PGM) \cite{pgm,Hall:2011jd}, which can be parametrized by a
single parameter \cite{eioy} $m_{3/2}$. This
minimal model of pure gravity mediation is similar in many ways to minimal supergravity
(mSUGRA). Universal masses equal to $m_{3/2}$ are imposed at the grand
unified theory (GUT) scale
based on the assumption that the K\"ahler manifold is flat for all matter
fields.  Unlike the CMSSM, gauginos do not get a tree-level
mass. This is because the supersymmetry breaking field is not a singlet
and so is excluded from coupling to the gauge kinetic function to leading
order. Thus, the leading order contribution to the mass of the gauginos
comes from anomaly mediation \cite{anom} and is loop suppressed relative to the
sfermion masses. The $B$-term, which contributes to electroweak symmetry
breaking, is identical to that in mSUGRA, $B=A-m_{3/2}$.  However, since
the $A$-terms of PGM are effectively zero, $B=-m_{3/2}$ and $B$ is fixed
for a given value of $m_{3/2}$. This makes radiative electroweak
symmetry breaking (EWSB) difficult. However, by adding a
Giudice-Masiero term \cite{gm}, $B$ is no longer
fixed by $m_{3/2}$ alone, but also depends on the coupling of the
Giudice-Masiero
term. This additional freedom in $B$ makes radiative EWSB possible, but
only for small values of $\tan\beta$.  Once the Higgs mass constraint is
taken into consideration, these models have a single free parameter which is
some combination of $m_{3/2}$ and $\tan\beta$ \cite{eioy}. However, because
$\tan\beta$ is restricted to be less than about 3, $m_{3/2}$ tends to be
rather large. The constraints on $\tan\beta$ can be removed, if the
Higgs soft masses at the GUT scale are taken to be non-universal
\cite{eioy2}. In this case, $m_{3/2}$ can be taken to be smaller for
larger values of $\tan\beta$.

Another important motivation for SUSY is grand unification
\cite{Georgi:1974sy}.  In the Standard Model (SM), the gauge couplings approach each
other as they are run up to the high scale \cite{Georgi:1974yf}.
However, the quality of the coupling unification is less than
convincing.  If the SM is supersymmetrized, on the other hand, the
unification of the gauge couplings becomes quite good
\cite{Dimopoulos:1981yj}. Furthermore,
grand unification in the SM would generate enormous quadratic
divergences for the Higgs boson. However, these quadratic divergences
are significantly reduced for supersymmetric grand unified theories, even
if the sfermions are larger than a TeV. Clearly, grand unification is
another motivation for PGM.

The signatures of these simple PGM-type models are limited. One possible
signature at the LHC for small $m_{3/2}$ is the wino \cite{winolhc}.  For larger
$m_{3/2}$, on the other hand, the wino cannot be seen at the LHC but
could be a viable thermal relic dark matter candidate
\cite{Hisano:2006nn}. If this is indeed the case, it could be seen by
indirect detection experiments in the near future
\cite{Bhattacherjee:2014dya}. However, this scenario is already under
tension from existing indirect detection experiments \cite{wino}. The
direct detection of wino dark matter is challenging as its scattering
cross section with
a nucleon is as small as $10^{-47}~\text{cm}^2$ \cite{winodd}. A
Higgsino signature at the LHC is another possible observable which
arises from tuning $\mu$ to be small  \cite{higgsinolhc}. However, this is also difficult to see. This scenario could also have Higgsino-like
dark matter which could possibly be seen in future indirect detection
experiments \cite{baer}. The scattering cross section of the Higgsino
with a nucleon is dependent on the size of the wino component of the
LSP, and may be probed in future experiments \cite{Hisano:2012wm}.

In this work, we will examine another possible signature of these
models. Since the colored triplet Higgs gives threshold corrections to
the gauge couplings when integrated out, the quality of the coupling
unification determines the mass of the colored triplet Higgs
\cite{Hisano:1992mh,Hisano:1992jj,Hisano:2013cqa} and so
affects the lifetime of the proton. When the colored Higgs is integrated
out, it also generates a dimension-five operator proportional to down-type
Yukawa couplings which lead to proton decay
\cite{dim5protondecay}. Since this dimension-five
operator is proportional to the down-type Yukawa couplings, it will be
enhanced for large $\tan\beta$. Proton decay from this dimension-five
operator arises from a loop diagram with a Higgsino mass
insertion \cite{higprotondec}. Proton decay of this type can then be
suppressed for small
$\mu$.  When unification is not ideal and $\tan\beta$ is large,  a
larger Higgsino mass can increase the rate of proton decay from this
dimension 5 operator. Parameters of this size are viable in PGM models.
Since proton decay of this type is also suppressed by $m_{3/2}$, the
more interesting parameter space will be for smaller $m_{3/2}$ and
larger $\tan\beta$. Therefore, we will need to consider non-universal
Higgs soft masses. We will find that if $m_{3/2}$ is small and
$\tan\beta$ is larger, which is also consistent with the Higgs mass
measurement, the proton lifetime may be in reach of future experiments.
However, for much of the parameter space the lifetime tends to be well
beyond the reach of future experiments. We will also look at the quality
of the gauge coupling unification determined by the deviation of the
colored triplet Higgs mass, $M_{H_C}$, from the GUT-scale as well as the
deviation of $\left(M_X^2M_\Sigma\right)^{1/3}$, where $X$ represents the
GUT scale SU(5) gauge bosons that become massive and $\Sigma$ is the ${\bf
24}$ which breaks SU(5) at the GUT scale.

\section{Minimal SUSY SU(5) GUT}
\label{sec:MSGUT}

In this section, we will outline the SU(5) SUSY GUT theory
\cite{Dimopoulos:1981zb, Sakai:1981gr} we will
consider.  Additional details on these models can be found in Appendix
\ref{App:NotCon}. The superpotential for this minimal SU(5) SUSY GUT is
given by
\begin{equation}
 W=W_{\text{Higgs}}+W_{\text{Yukawa}} ~,
\label{eq:superpotential}
\end{equation}
where
\begin{align}
 W_{\rm Higgs} &= \frac{1}{3}\lambda_{\Sigma}{\rm Tr}\Sigma^3
 +\frac{1}{2}m_\Sigma
 {\rm Tr} \Sigma^2 +\lambda_H \bar{H}\Sigma H +m_H\bar{H} H ~,
\label{superpotentialHiggs}
\\
 W_{\rm Yukawa} &=
\frac{1}{4}h^{ij}\epsilon_{\hat{a}\hat{b}\hat{c}\hat{d}\hat{e}}\Psi_i^{\hat{a}
 \hat{b}} \Psi_j^{\hat{c}\hat{d}}H^{\hat{e}} -\sqrt{2}
f^{ij}\Psi_i^{\hat{a}\hat{b}} \Phi_{j\hat{a}}\bar{H}_{\hat{b}}~,
\label{superpotentialYukawa}
\end{align}
and $\hat{a},\hat{b},\dots=1$--$5$ represent the SU(5) indices and
$\epsilon_{\hat{a}\hat{b}\hat{c}\hat{d}\hat{e}}$ is the totally
antisymmetric tensor with $\epsilon_{12345}=1$.  $\Phi_i$ and $\Psi_i$
are the chiral superfields in the $\bar{\bf 5}$ and ${\bf 10}$ representations,
respectively, with $i$ denoting the generation index. $H$ and $\bar H$
are the ${\bf 5}$ and $\bar {\bf 5}$ containing the mininal
supersymmetric Standard Model (MSSM) doublets. In
these expressions, we have assumed $R$-parity conservation which forbids
terms like $\Psi\Phi\Phi$ and $H\Phi$. The adjoint Higgs field,
$\Sigma$, gets a vacuum expectation value (VEV) in the direction
\begin{equation}
 \langle \Sigma \rangle =V\cdot {\rm diag}(2,2,2,-3,-3)~,
\end{equation}
breaking the SU(5) gauge group to the SM gauge groups
SU(3)$_C\otimes$SU(2)$_{L}\otimes$U(1)$_Y$. Because SUSY remains
unbroken for SU(5) breaking, we have $V=m_\Sigma/\lambda_\Sigma$. For
this setup, the masses of $\Sigma_3$, $\Sigma_8$, $\Sigma_{24}$, and
$H_C$ are given as
\begin{equation}
 M_{\Sigma}\equiv
M_{\Sigma_8}=M_{\Sigma_3}=\frac{5}{2}\lambda_{\Sigma}V~,
~~~~~
M_{\Sigma_{24}}=\frac{1}{2}\lambda_{\Sigma}V~,
~~~~~
M_{H_C}=5\lambda_H V~,
\end{equation}
while the $\mu$ term for the MSSM Higgs fields is
\begin{equation}
 \mu_0 =m_H -3\lambda_H V ~.
\end{equation}
As is usually done, we tune the parameter $m_H$  to
realize $\mu_0\ll m_H$ which is typically referred to as the
doublet-triplet splitting.\footnote{This is another fine-tuning
besides that for the Higgs mass. Note that the fine-tuning for the Higgs
mass becomes worse as the $\mu$ parameter is taken to be larger, while the
doublet-triplet fine-tuning becomes less severe. This tension may
explain why $\mu$ is much larger than the electroweak scale
\cite{Nomura:2014asa, Yanagida:2014}.} In addition, the gauge
interactions of the adjoint Higgs field yield an $X$-boson mass of
$M_X=5\sqrt{2}g_5 V$ where $g_5$ is the unified gauge coupling
constant. The
components $\Sigma_{(3^*, 2)}$ and $\Sigma_{(3,2)}$ become the
longitudinal component of the $X$ bosons, and thus do not appear as
physical states.

The Yukawa couplings $h^{ij}$ and $f^{ij}$ in
Eq.~(\ref{superpotentialYukawa}) have redundant degrees of freedom, most
of which are eliminated by the field redefinition of $\Psi$ and $\Phi$.
Since $h^{ij}$ is a symmetric matrix, $h^{ij}$ and $f^{ij}$ have
six and nine complex degrees of freedom, respectively. The field
redefinition of the SM fields forms the U(3)$\otimes$U(3) transformation
group, and thus the physical degrees of freedom turn out to be
$(12+18)-9\times 2=12$. Among these degrees of freedom, six of them are
the quark mass eigenvalues and four are for the
Cabibbo-Kobayashi-Maskawa (CKM) matrix elements and we are left with two
phases \cite{Ellis:1979hy}. In this paper, we take the same basis used
in Ref.~\cite{Hisano:1992jj} such that 
\begin{align}
 h^{ij}&= e^{i\varphi_i} \delta_{ij}f_{u_i}(Q_{{G}}) ~, \\
 f^{ij}&= V^*_{ij} f_{d_j}(Q_{{G}}) ~,
\end{align}
where $f_{u_i}(Q_{{G}})$ and $f_{d_j}(Q_{{G}})$ are
the up-type and down-type Yukawa couplings, respectively, at a scale
$Q_{{G}}$ around the GUT scale, and $V_{ij}$ is the
CKM matrix. The phase factors $\varphi_i$
satisfy the condition $\sum_{i}\varphi_i=0$, and thus only two of them are
independent.
In this basis, the MSSM superfields are embedded into the SU(5) matter
multiplets as
\begin{align}
 \Psi_i&\ni \{
Q_i,~e^{-i\varphi_i}\overline{U}_i,~V_{ij}\overline{E}_j
\}~,~~~~~~
\Phi_i\ni\{
\overline{D}_i,~L_i
\}~.
\end{align}
Then, Eq.~\eqref{superpotentialYukawa} leads to
\begin{align}
 W_{\rm Yukawa}&=f_{u_i}
(Q^{a}_i\cdot H_2)\overline{U}_{ia}-V^*_{ij}f_{d_j} (Q^{a}_i\cdot H_1)
\overline{D}_{ja}-f_{d_i}
\overline{E}_i (L_i\cdot H_1)\nonumber \\[2pt]
&-\frac{1}{2}e^{i\varphi_i}\epsilon_{abc}f_{u_i}
(Q^{a}_i \cdot Q^{b}_i) H^c_C
+V^*_{ij}f_{d_j}(Q^{a}_i\cdot L_j)\overline{H}_{Ca}
\nonumber \\[2pt]
&+f_{u_i}V_{ij}\overline{U}_{ia}
\overline{E}_jH^a_C
-V^*_{ij}f_{d_j}e^{-i\varphi_i}\epsilon^{abc}
\overline{U}_{ia}\overline{D}_{jb}\overline{H}_{Cc}~.
\label{eq:wyukawa}
\end{align}
The new phase factors appear only in the couplings of the color-triplet
Higgs multiplets.

\section{Mass Spectrum and Coupling Unification}

To compute the proton decay rate, we need to evaluate the masses of the
GUT-scale particles which induce the baryon-number violating
interactions. In this section, we estimate these masses using the method
discussed in Refs.~\cite{Hisano:1992mh,Hisano:1992jj,Hisano:2013cqa}.
The mass of the
heavy particles is determined by first RG running the couplings to the
scale where they approximately unify. Then, because the thresholds at
the GUT scale depend on these superheavy particles, their masses can be
determined by assuming the deviation in gauge coupling unification is
solely due to these thresholds. Note, we will use the $\overline{\rm
DR}$ scheme \cite{Siegel:1979wq} in the following calculation. At the
scale $Q_G$ near the GUT scale, the one-loop matching conditions for
the gauge coupling constants are as follows
\cite{Weinberg:1980wa, Hall:1980kf}:
\begin{align}
 \frac{1}{g_1^2(Q_{G})}&=\frac{1}{g_G^2(Q_G)}
+\frac{1}{8\pi^2}\biggl[
\frac{2}{5}
\ln \frac{Q_{G}}{M_{H_C}}-10\ln\frac{Q_{G}}{M_X}
\biggr]~,\nonumber \\
 \frac{1}{g_2^2(Q_{G})}&=\frac{1}{g_G^2(Q_{G})}
+\frac{1}{8\pi^2}\biggl[
2\ln \frac{Q_{G}}{M_\Sigma}-6\ln\frac{Q_{G}}{M_X}
\biggr]~,\nonumber \\
 \frac{1}{g_3^2(Q_{G})}&=\frac{1}{g_G^2(Q_{G})}
+\frac{1}{8\pi^2}\biggl[
\ln \frac{Q_{G}}{M_{H_C}}+3\ln \frac{Q_{G}}
{M_\Sigma}-4\ln\frac{Q_{G}}{M_X}
\biggr]~,
\end{align}
where $g_G$ is the unified gauge coupling constant.
Note that the conditions do not include constant (scale independent) terms
since we use the $\overline{\rm DR}$ scheme for
renormalization. Assuming the above equations contain the major
thresholds for the gauge couplings, they can be used to solve for the
masses
\begin{align}
 \frac{3}{g_2^2(Q_{G})}- \frac{2}{g_3^2(Q_{G})}
- \frac{1}{g_1^2(Q_G)}
&=-\frac{3}{10\pi^2}\ln \biggl(\frac{Q_{G}}{M_{H_C}}\biggr)
~, \nonumber \\
 \frac{5}{g_1^2(Q_{G})}- \frac{3}{g_2^2(Q_{G})}
- \frac{2}{g_3^2(Q_{G})}
&=-\frac{9}{2\pi^2}\ln\biggl(
 \frac{Q_{G}}{M_{G}}\biggr)~,
\label{conditions}
\end{align}
with $M_{\text{G}}\equiv (M_X^2M_{\Sigma})^{\frac{1}{3}}$.
The above expressions allow us to find the masses of the heavy particles
in the combination\footnote{The third condition is used to determine
$g_G^2(Q_G)$.}, $M_{H_C}$ and $M_X^2M_{\Sigma}$. The value of $M_{H_C}$
and $M_G$ found from these relationships will be used below to find the
lifetime of the proton.

\section{Proton Decay}

In the
minimal SUSY GUT, proton decay is induced by the exchange of the
color-triplet Higgs boson, and the dominant decay mode is, generally,
$p\to K^+\bar{\nu}$ \cite{dim5protondecay}. We will only
give details of the contributions from the colored Higgs boson since it
will often be the dominant source of proton decay in PGM. At the GUT scale,
the triplet Higgs boson is integrated out. The most important
interaction for our considerations is the color-triplet Higgs exchange
which we match at the scale $Q_G$ on to the dimension-five effective
Lagrangian
\begin{equation}
{\cal L}_5^{\rm eff}= C^{ijkl}_{5L}{\cal O}^{5L}_{ijkl}
+C^{ijkl}_{5R}{\cal O}^{5R}_{ijkl}
~~+~~{\rm h.c.}~,
\end{equation}
where the effective operators ${\cal O}^{5L}_{ijkl}$ and ${\cal
O}^{5R}_{ijkl}$ are defined by
\begin{align}
 {\cal O}^{5L}_{ijkl}&\equiv\int d^2\theta~ \frac{1}{2}\epsilon_{abc}
(Q^a_i\cdot Q^b_j)(Q_k^c\cdot L_l)~,\nonumber \\
{\cal O}^{5R}_{ijkl}&\equiv\int d^2\theta~
\epsilon^{abc}\overline{U}_{ia}\overline{E}_j\overline{U}_{kb}
\overline{D}_{lc}~,
\label{eq:dim5operators}
\end{align}
and the Wilson coefficients $C^{ijkl}_{5L}$ and $C^{ijkl}_{5R}$ are
given by
\begin{align}
 C^{ijkl}_{5L}(Q_G)&
=\frac{1}{M_{H_C}}f_{u_i}
e^{i\varphi_i}\delta^{ij}V^*_{kl}f_{d_l}~,\nonumber \\
C^{ijkl}_{5R}(Q_G)
&=\frac{1}{M_{H_C}}f_{u_i}V_{ij}V^*_{kl}f_{d_l}
e^{-i\varphi_k}
~.
\label{eq:wilson5}
\end{align}
Note, the color indices must be completely antisymmetric for these
interactions and as a result,
only operators with at least two generations will be allowed. For this
reason, the dominant decay modes contain a strange quark in their final
state, \textit{i.e.}, $p\to K^+\bar{\nu}$.

As can be seen in Eq.~\eqref{eq:wyukawa}, at the GUT scale the lepton
and down-type quark Yukawa couplings should be equal.  However, in
running up from the weak scale, we find them to be quite different
especially those for the first two generations. The difference is,
however, easily compensated by effects above the GUT scale; for
instance, the higher-dimensional operators induced at the Planck scale
contribute to the Yukawa couplings, which may account
for this difference \cite{Nath:1996qs, Nath:1996ft,
Bajc:2002pg}. Because it is not known which of these values is
close to the correct value for the Yukawa coupling at the GUT scale, in
the the discussion below, we use both the down quark and lepton-type
Yukawa couplings to calculate the proton lifetime. This will allow us to
quantify our uncertainty in the lifetime of the proton.

The relevant operators in Eq.~\eqref{eq:wilson5} can be further reduced
by keeping only those with the largest Yukawa couplings.  We find that
only the operators ${\cal O}^{5R}_{3312}$ and ${\cal O}^{5R}_{3311}$
yield a sizable contribution to proton decay, even though the contribution
is suppressed by a flavor changing element of the CKM matrix. This
contribution turns out to be dominant because of the large third
generation Yukawa couplings involved \cite{higprotondec}. The relevant
Wilson coefficients are then
\begin{align}
 C^{3311}_{5R}
(Q_G)=\frac{1}{M_{H_C}}f_tf_d(Q_G)
e^{-i\varphi_1}V_{tb}V_{ud}^*~, \nonumber \\
 C^{3312}_{5R}(Q_G)=\frac{1}{M_{H_C}}f_tf_s(Q_G)
e^{-i\varphi_1}V_{tb}V_{us}^*~.\label{eq:wilson3genT}
\end{align}
Notice that the coefficients include a common phase factor
$e^{-i\varphi_1}$, which is therefore not important for proton decay.

The Wilson coefficients in Eq.~\eqref{eq:wilson3genT} are then evolved down to
the SUSY scale. At the SUSY scale, the sfermions of
these dimension-five operators are integrated out
via the one-loop diagram found in Fig.~\ref{fig:1loop} of
Appendix~\ref{App:ProDec}.  The process proceeds via the exchange of
either a  charged wino or a Higgsino.\footnote{This is the dominant
contribution to proton decay, unless there is flavor violation in the sfermion
sector. In this paper, we assume there is no flavor violation in the
sfermion sector. The flavor violating case is discussed in
Ref.~\cite{Nagata:2013sba}.} In PGM, we generally have
$|\mu|\gg |M_2|$ and so the contribution from Higgsino exchange
dominates \cite{Hisano:2013exa}.\footnote{ Higgsino exchange dominates
in this limit because the gauginos and Higgsinos in the one-loop
diagrams are required to flip their chirality, and thus their
contribution to proton decay is proportional to their masses, as can be
seen from the expression for the function $F$ given in
Eq.~\eqref{eq:funceq}. } For these reasons, we focus on the charged
Higgsino exchange process in what follows.

The loop diagram in Fig.~\ref{fig:1loop} is then matched onto the
baryon-number violating four-fermion operators \cite{Weinberg:1979sa,
Wilczek:1979hc, Abbott:1980zj}
\begin{equation}
 {\cal L}^{\text{eff}}_6=C_i~ \epsilon_{abc}(u^a_{R1}d^b_{Ri})
(Q_{L3}^c \cdot L_{L3})~,
\end{equation}
with
\begin{equation}
 C_i (Q_S)=\frac{f_tf_\tau}{(4\pi)^2}C^{*331i}_{5R}(Q_S)
F(\mu, m_{\widetilde{t}_R}^2,m_{\tau_R}^2)~,
\label{eq:susymatching}
\end{equation}
where $i=1,2$, and $Q_S$ is the SUSY breaking scale taken to be around
$m_{3/2}$. The loop function $F$ is found in Appendix
\ref{App:ProDec}. The above expression shows that the proton decay rate
depends on the SUSY spectra through the loop function. We will see this
dependence in Sec.~\ref{sec:results} for the PGM scenario. Note that the
loop function is suppressed by the sfermion masses. Thus, we expect that
for large $m_{3/2}$ the proton lifetime is long enough
\cite{Hisano:2013exa, Liu:2013ula} to evade the
current bound, $\tau (p \to K^+ \bar{\nu}) > 5.9\times 10^{33}$~years
\cite{Abe:2014mwa}. This can be compared to the weak-scale SUSY
scenarios; in these cases, the proton decay rate is in general predicted
to be so large that the minimal SUSY GUT is excluded
\cite{Murayama:2001ur} and thus some additional conspiracy is required
to realize a SUSY GUT.

We now run the Wilson coefficients down to the hadronic scale,
$Q_{\text{had}}=2$ GeV. The Lagrangian at this scale takes the
form\footnote{For more details of how we arrived at this expression see
Appendix~\ref{App:ProDec}.}
\begin{equation}
 {\cal L}(p\to K^+\bar{\nu}_\tau)
=C_{usd} [\epsilon_{abc}(u_R^as_R^b)(d_L^c\nu_\tau)]
+C_{uds} [\epsilon_{abc}(u_R^ad_R^b)(s_L^c\nu_\tau)]
~.
\end{equation}
Using these Wilson coefficients, we then evaluate
the partial decay width of the $p\to K^+ \bar{\nu}$ and find
\begin{equation}
 \Gamma(p\to K^+\bar{\nu})=\frac{m_p}{32\pi}
\biggl(1-\frac{m_K^2}{m_p^2}\biggr)^2
|{\cal A}(p\to K^+\bar{\nu})|^2~,
\label{tp5}
\end{equation}
where $m_p$ and $m_K$ are the masses of proton and kaon, respectively,
and
\begin{equation}
 {\cal A}(p\to K^+\bar{\nu})=
C_{usd}(Q_{\text{had}})\langle K^+\vert (us)_R ^{}d_L^{}\vert p\rangle +
C_{uds}(Q_{\text{had}})\langle K^+\vert (ud)_R ^{}s_L^{}\vert p\rangle ~.
\end{equation}
The hadron matrix elements in the above equation have been recently
computed in Ref.~\cite{Aoki:2013yxa} using a lattice simulation of QCD,
\begin{align}
 \langle K^+\vert (us)_R ^{}d_L^{}\vert p\rangle&=
-0.054(11)(9)~\text{GeV}^2
~,\nonumber\\
\langle K^+\vert (ud)_R ^{}s_L^{}\vert p\rangle &=
-0.093(24)(18)~\text{GeV}^2
~,
\end{align}
where the first and second parentheses represent statistical and
systematic errors, respectively. The matrix elements are computed at the
scale $Q_{\text{had}}=2$~GeV.

Before concluding this section, we comment on other possible
contributions to proton decay. Firstly, the dimension-five baryon-number
violating operators in Eq.~\eqref{eq:dim5operators} can also be generated at
the Planck scale, $M_P$. If the coefficients of the operators are
${\cal O}(1/M_P)$, that is,  there is no suppression from Yukawa couplings, then they
will give the dominant contribution to proton decay and result in a
lifetime which is too short \cite{Dine:2013nga}. It is expected, however, that there is some underlying mechanism such as a flavor
symmetry which is responsible for the structure of the Yukawa couplings. This symmetry could give additional suppression to these Planck-scale
operators. In this paper, we assume that the contribution
of these operators is less significant compared with the color Higgs
contribution, and neglect them in the following analysis.

Secondly, the exchange of the $X$ bosons will also induce proton
decay. This decay mode is via a dimension-six GUT-scale effective
operator and is thus usually subdominant compared to the contribution
of the dimension-five operator discussed above. An approximate
expression for the lifetime of the proton from the dimension-six
operator is
\begin{eqnarray}
\tau(p\to e^+\pi^0) \simeq 3 \times 10^{35} \times
 \left(\frac{M_X}{1.0\times 10^{16}~\text{GeV}}\right)^4 ~.
 \label{tp6}
\end{eqnarray}
There is a slight dependence on the masses of SUSY particles we have
neglected.  As can be seen from this expression,
the proton decay width from the dimension-six operator will in general
give lifetimes too long to be detected, at least much longer than the
present bound: $\tau (p\to e^+\pi^0)> 1.4 \times 10^{34}~{\rm years}$
\cite{Shiozawa, Babu:2013jba}.

\section{Pure Gravity Mediation}
As discussed above, the lifetime of the proton depends on the SUSY
parameters.  Motivated by the $126$ GeV Higgs boson \cite{lhch} and other
cosmological considerations \cite{ego}, we will analyze the proton lifetime for PGM
models. The scalar potential of PGM takes the same form as that of
mSUGRA
 \begin{eqnarray}
V  & =  &  \left|{\partial W \over \partial \phi^i}\right|^2 +
\left( A_0 W^{(3)} + B_0 W^{(2)} + \text{h.c.}\right)  + m_{3/2}^2
\phi^i \phi_i^*  \, ,
\end{eqnarray}
which is determined by the flat K\"ahler manifold\footnote{If the
K\"ahler manifold for the first two generations is no-scale like, these
models can explain $g-2$ experiments \cite{Evans:2013uza}.  However, in
this case the proton decay calculation is more complicated because of
an additional wino contribution but should give a similar order of
magnitude for the proton lifetime.} and the superpotential $W$ is given
in Eq.~\eqref{eq:superpotential}. $W^{(2)}$ and $W^{(3)}$ are the bi-
and trilinear parts of the superpotential. 
For PGM, the SUSY breaking field is a non-singlet and strongly
stabilized \cite{strongStab} which suppresses the gaugino masses and
$A$-terms respectively. The gaugino masses are regenerated by anomalies
and take the form\footnote{The $A$-terms are also regenerated by
anomalies. However, they are too small to be of importance.} 
\begin{eqnarray}
    M_{1} &=&
    \frac{33}{5} \frac{g_{1}^{2}}{16 \pi^{2}}
    m_{3/2}\ ,
    \label{eq:M1} \\
    M_{2} &=&
    \frac{g_{2}^{2}}{16 \pi^{2}} m_{3/2}  \ ,
        \label{eq:M2}     \\
    M_{3} &=&  -3 \frac{g_3^2}{16\pi^2} m_{3/2}\ .
    \label{eq:M3}
\end{eqnarray}
In order to account for radiative EWSB,
mSUGRA is further modified by including a Giudice-Masiero term for the Higgs fields in the K\"ahler manifold
\cite{gm}.  This modifies the Higgs boson parameters to
\begin{eqnarray}
 \mu &=& \mu_0 + c_H m_{3/2}\ ,
 \label{eq:mu0}
 \\
  B\mu &=&  \mu_0 (A_0 - m_{3/2}) + 2 c_H m_{3/2}^2\ ,
   \label{eq:Bmu0}
\end{eqnarray}
where $\mu_0$ is the superpotential Higgs bilinear term found in $W^{(2)}$.
This allows us to vary both $\mu$ and $B\mu$ independently in order to
satisfy the EWSB conditions. This leaves $m_{3/2}$ and $c_H$ as free
parameters. In this case, $\tan \beta$ is an output of the EWSB conditions,
but in practice one can trade $c_H$ for $\tan \beta$ and use $m_{3/2}$
and $\tan \beta$ as free inputs.   Since this simplest of PGM models
tends to require small $\tan\beta$ and larger $m_{3/2}$, we will allow
the Higgs soft masses to be free parameters. This will allow $\tan\beta$
to be larger and so allow for $m_{3/2}$ to be smaller \cite{eioy2}. As
was seen in the previous sections, both larger $\tan \beta$ and smaller
$m_{3/2}$ will lead to shorter lifetimes of the proton.  We will not
discuss the origin of these non-universal Higgs soft masses here.
However, discussion about this can be found in Ref.~\cite{eioy2}. Lastly,
we note that the non-universal Higgs soft masses, $m_1$ and $m_2$, can
also be parametrized in terms of the low scale values of $\mu$ and
$m_A$ which are otherwise also outputs of the EWSB conditions. We will
take advantage of this in the results below in order to zoom in on some
features of the proton lifetime.

\section{Results}
\label{sec:results}

\begin{figure}[t!]
\scalebox{0.5}{\includegraphics{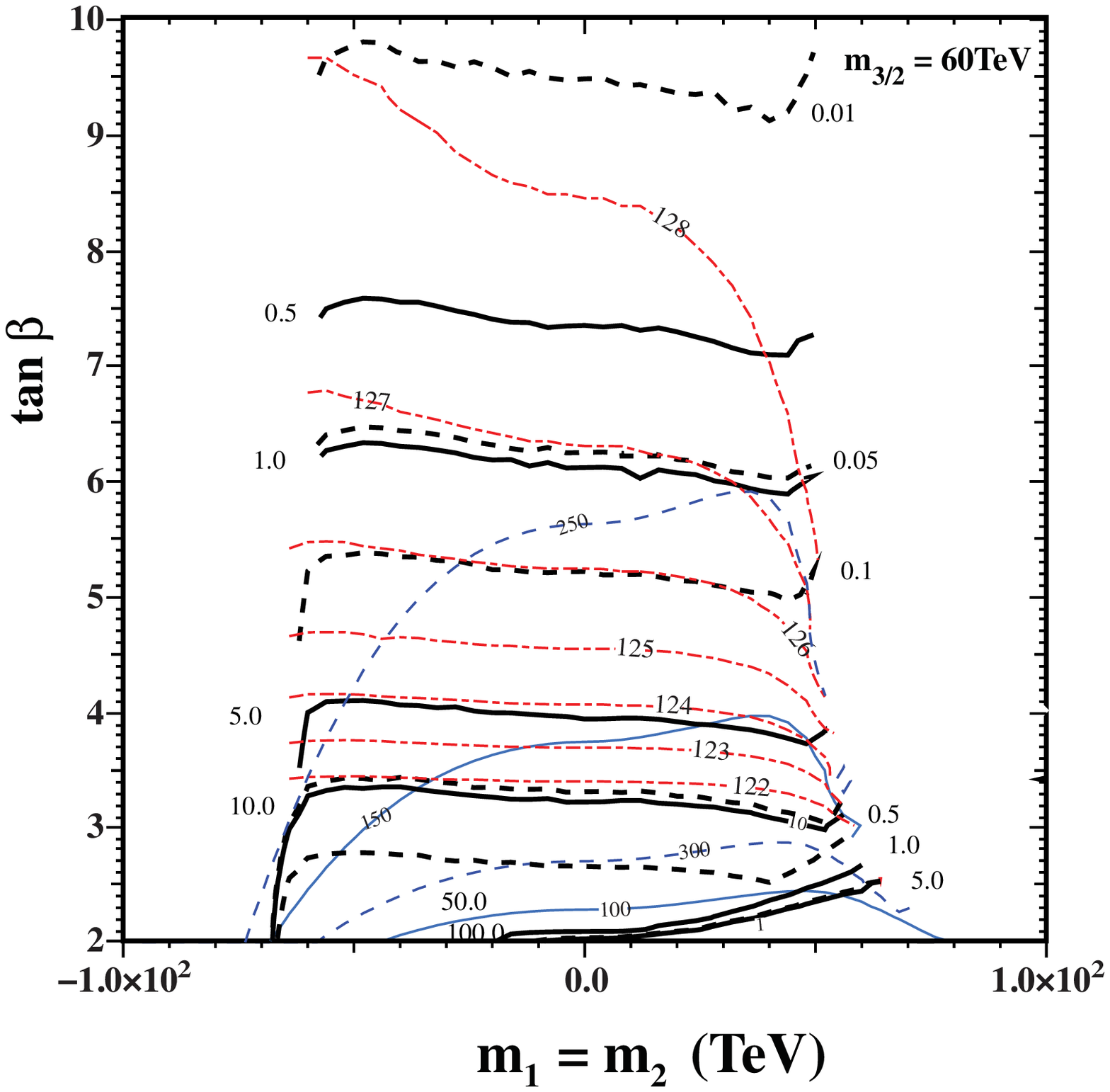}}
\scalebox{0.5}{\includegraphics{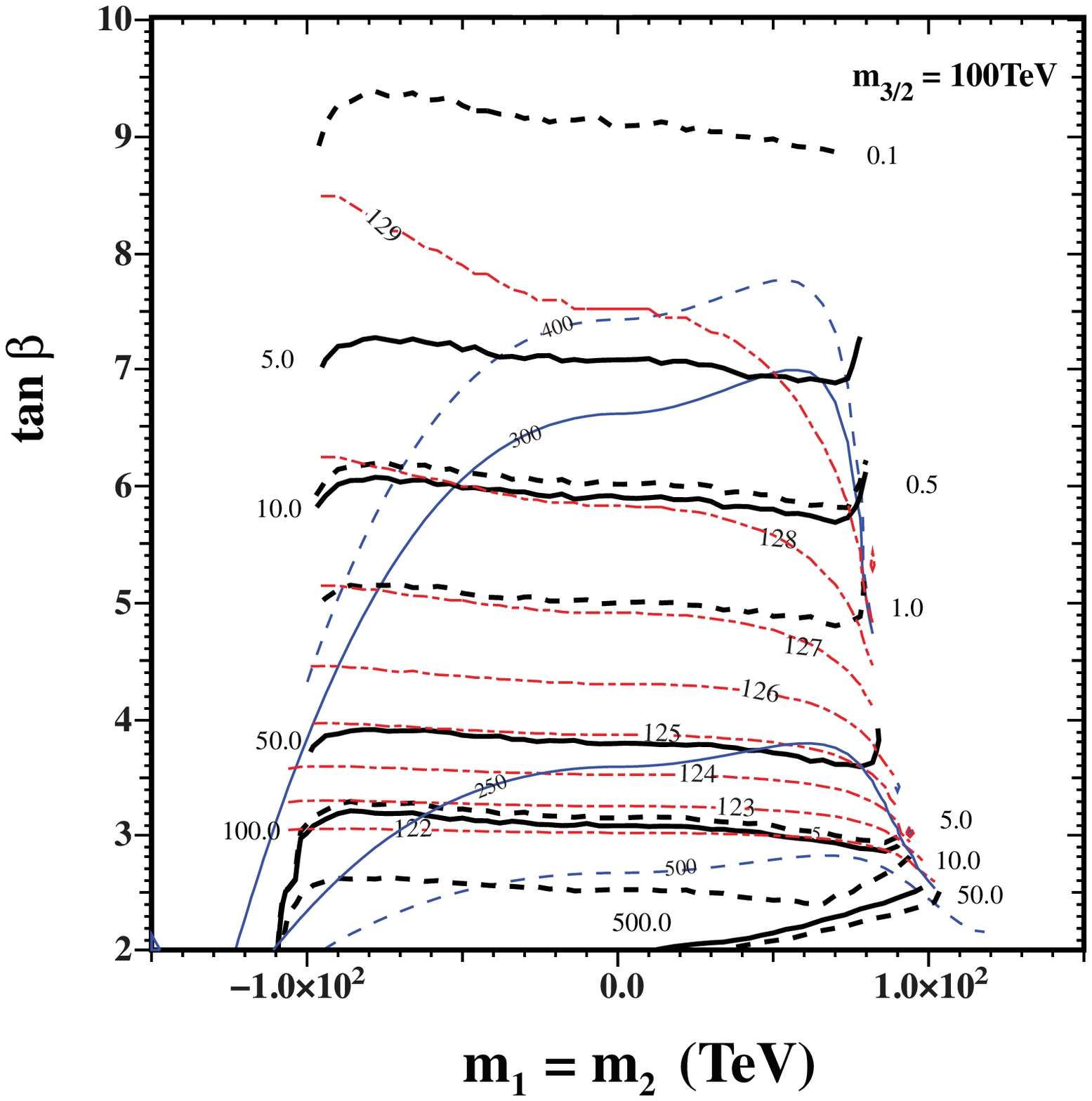}}
\caption{
{\it
The $\tan \beta$--$m_{1,2}$ plane for a) $m_{3/2} = 60$\,TeV and b)   $m_{3/2} = 100$\,TeV.
The Higgs mass is shown by the nearly horizontal thin red contours in $1$\,GeV intervals.
The wino/chargino mass is shown by the thin solid ($\mu > 0$) and dashed
($\mu < 0$) contours. The thick black contours show the value of the proton lifetime
based on the quark Yukawa couplings (solid) and lepton Yukawa couplings (dashed) in
units of $10^{35}$ years. Lifetime contours for the solid curves are labeled to the left of the contours
whereas dashed contours are labeled to the right.}}
\label{m1eqm2p}
\end{figure}

We are now in a position to discuss the proton lifetime and mass scales
associated with gauge coupling unification in a variety of models which have
varying degrees of non-universality in the Higgs sector.
We begin by displaying in Fig.~\ref{m1eqm2p} the $m_1=m_2$ vs. $\tan \beta$ plane
for fixed gravitino mass.
This is a one-parameter extension of the minimal (two-parameter) PGM model and resembles NUHM1 models \cite{nuhm1}.  In the left panel we have fixed $m_{3/2} = 60$ TeV.
For this value of the gravitino mass, the Higgs mass lies between 124 and 128 GeV
\footnote{We refer to this extended range of Higgs masses to account for the uncertainty in
the calculation of the Higgs mass. Note also that the Higgs masses calculated here
differ slightly from those calculated in \cite{eioy2} as here we are not imposing strict gauge coupling unification
at the GUT scale.} for $\tan \beta$ roughly between 4--9 as shown by the red dot-dashed curves.
The thin blue lines show the values of the LSP (wino) mass\footnote{The
present lower limit on the wino mass from the LHC experiment is about
270~GeV \cite{Aad:2013yna}. } and are solid for $\mu > 0$ and dashed for $\mu < 0$.  The anomaly mediated contribution to $m_\chi$  for $m_{3/2} = 60$ TeV
is 170 GeV. At low $\tan \beta$, threshold corrections from the heavy Higgs bosons and the Higgsinos increase the mass for $\mu < 0$ and decrease the mass for $\mu > 0$. At large $\tan \beta$, the wino mass,
for both positive and negative $\mu$, tends to its anomaly mediated value.
The curves end at high and low values of the Higgs soft masses due to the absence of radiative electroweak
symmetry breaking. For large and negative values of $m_1^2 = m_2^2$ (the sign on the axis refers to the sign
of the mass squared), the Higgs pseudoscalar mass squared is negative, and for large positive
$m_1^2 = m_2^2$, the electroweak conditions yield $|\mu|^2 < 0$.

The thicker black curves in Fig.~\ref{m1eqm2p} show the values of the proton lifetime. As discussed earlier,
as there is some uncertainty as to how we match the Yukawa couplings at the GUT scale,
we have results based on quark Yukawa couplings (shown by the solid curves) and results based on
lepton Yukawa couplings (shown by the dashed curves). As one can see from the figure,
the calculated proton lifetime is sensitive to $\tan \beta$ yet relatively insensitive to the value of $m_{1,2}$ for fixed
gravitino mass.  In general, the proton lifetime is lower at high $\tan \beta$ due to the increase in the down-like Yukawa couplings when $\tan \beta$ is increased, whereas the Higgs mass increases
with $\tan \beta$.
For these relatively low values of the gravitino mass used in the left panel,
the proton lifetimes based on quark Yukawas drop below $5 \times 10^{34}$ years only when
 $\tan \beta > 7$ where $m_h > 127$ GeV. The lifetime increases rapidly at lower $\tan \beta$ and
 exceeds $5 \times 10^{35}$ years when $\tan \beta < 4$ where $m_h < 124$ GeV. However, the wino mass requires $\mu<0$ and $\tan\beta\gtrsim 6$.
 Recall that these lifetimes are computed from Eq.~(\ref{tp5}) and when the lifetime exceeds
 3 $\times 10^{35}$ years, the dominant contribution to the decay rate comes from the
 dimension-six operator given in Eq.~(\ref{tp6}).
 Proton lifetimes based on lepton Yukawas are significantly smaller (by a factor of roughly 20),
 so that $\tau_p^l < 5 \times 10^{33}$ years when $\tan \beta \gtrsim 6$ and is still smaller than
 $2 \times 10^{34}$ years when $\tan \beta > 4$.

 In the right panel of Fig.~\ref{m1eqm2p}, we have taken $m_{3/2} = 100$ TeV and as expected
 the Higgs mass for a given value of $\tan \beta$ is higher.  The range 124 -- 128 GeV now requires
 $\tan \beta \simeq 3.5$ -- 6. The uncorrected wino mass is now about 290 GeV and in the figure
 we see lower (higher) wino masses when $\mu > (<) 0$. The proton lifetimes are now significantly higher.
 At $\tan \beta = 6$, the quark based value of $\tau_p$ determined by the dimension five operator is now $10^{36}$ years and increases as $\tan \beta$ is lower.  The lepton based lifetimes remain a factor of about 20 lower and may still be as low as $5 \times 10^{34}$ years
 at $\tan \beta = 6$.

\begin{figure}[t!]
\scalebox{0.6}{\includegraphics{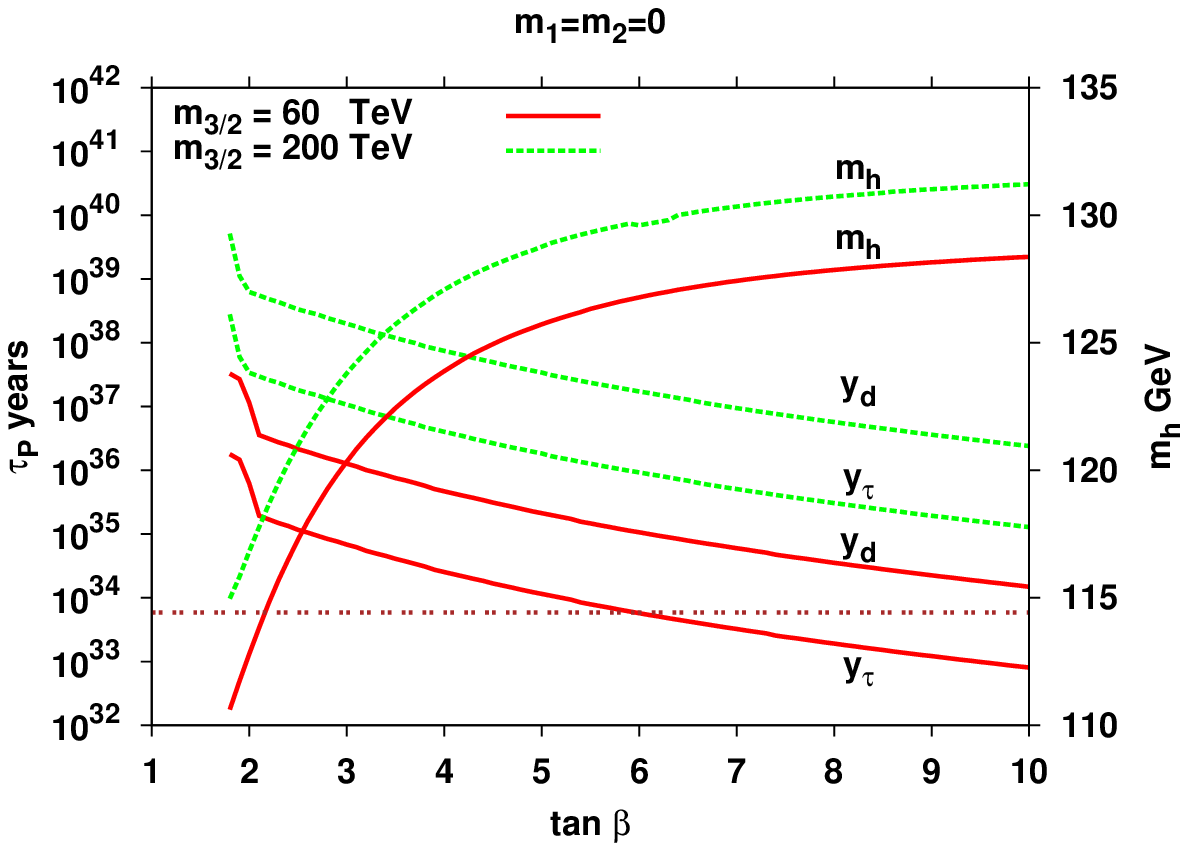}}
\scalebox{0.6}{\includegraphics{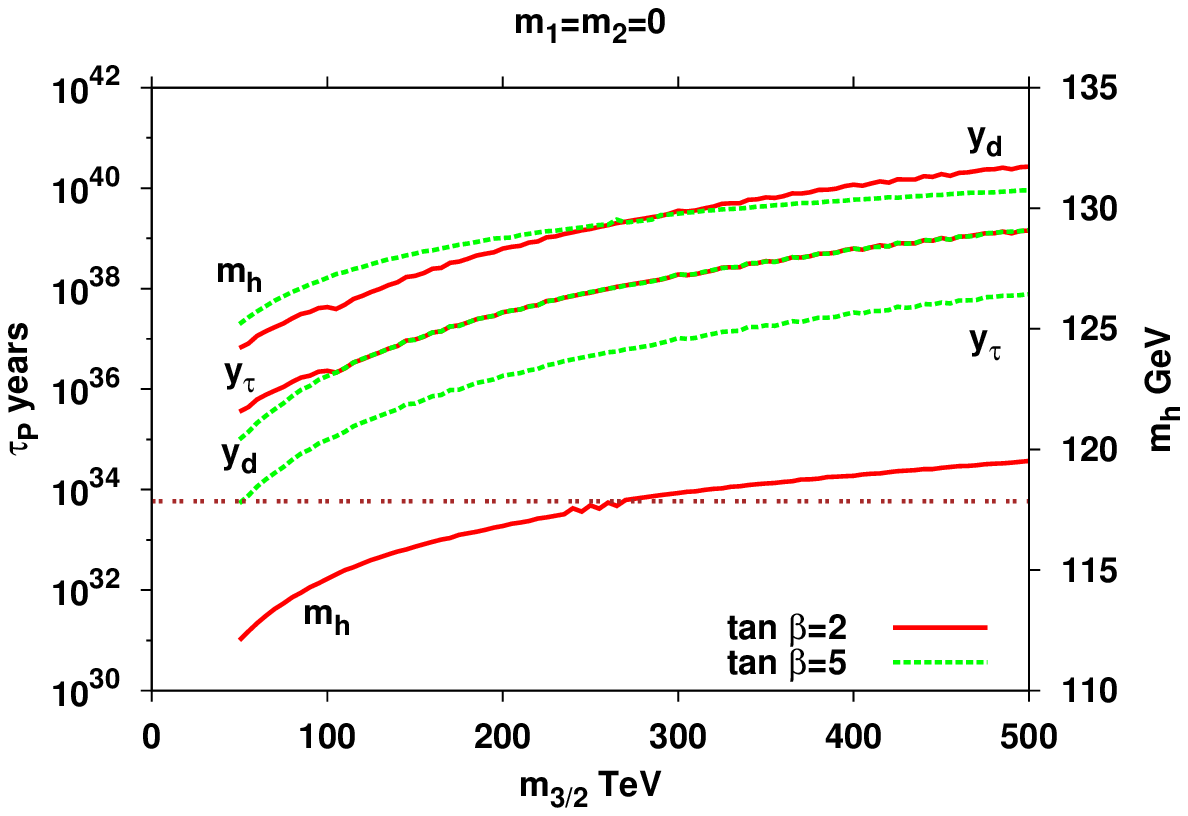}}
\caption{
{\it
The dependence of the proton lifetime on $\tan \beta$ (left) for fixed
 $m_{3/2} = 60$ and $200$ TeV, and on $m_{3/2}$ (right) for fixed $\tan \beta = 2$ and 5. Both of the Higgs soft masses have been fixed $m_1 = m_2 = 0$.
The dependence of the Higgs mass is also shown with its value given on
 the right side of each panel. The horizontal line indicates the present
 experimental bound \cite{Abe:2014mwa}.}}
\label{tpm32tb}
\end{figure}

To see more clearly the dependence of the proton lifetime on the PGM parameters,
we show in the left panel of Fig.~\ref{tpm32tb} the behavior of the proton
lifetime as a function of $\tan \beta$ for fixed $m_{3/2} = 60$ and 200
TeV with $m_1 = m_2 = 0$. The lifetime falls off monotonically with
$\tan \beta$. The ratio between the quark and lepton evaluation of
$\tau_p$ is seen to be nearly constant as $\tan \beta$ is varied with a
ratio of about 20. We also see the substantial increase in $\tau_p$ when
$m_{3/2}$ is increased to 200 TeV. The experimental limit on the proton
lifetime, $\tau (p \to K^+ \bar{\nu}) > 5.9\times 10^{33}$~years
\cite{Abe:2014mwa}, is shown by the horizontal line. In the right panel, we
show this increase in $\tau_p$ with $m_{3/2}$ for fixed values of $\tan
\beta = 2$ and 5. In both panels, we also show the Higgs mass as a
function of $\tan \beta$ and $m_{3/2}$ with its value displayed on the
right edge of each panel.  Restricting the Higgs mass to the range
124--128 GeV allows one to focus on the relevant ranges of either $\tan
\beta$ or $m_{3/2}$ and hence on the predicted proton lifetime.

\begin{figure}[t!]
\scalebox{0.63}{\includegraphics{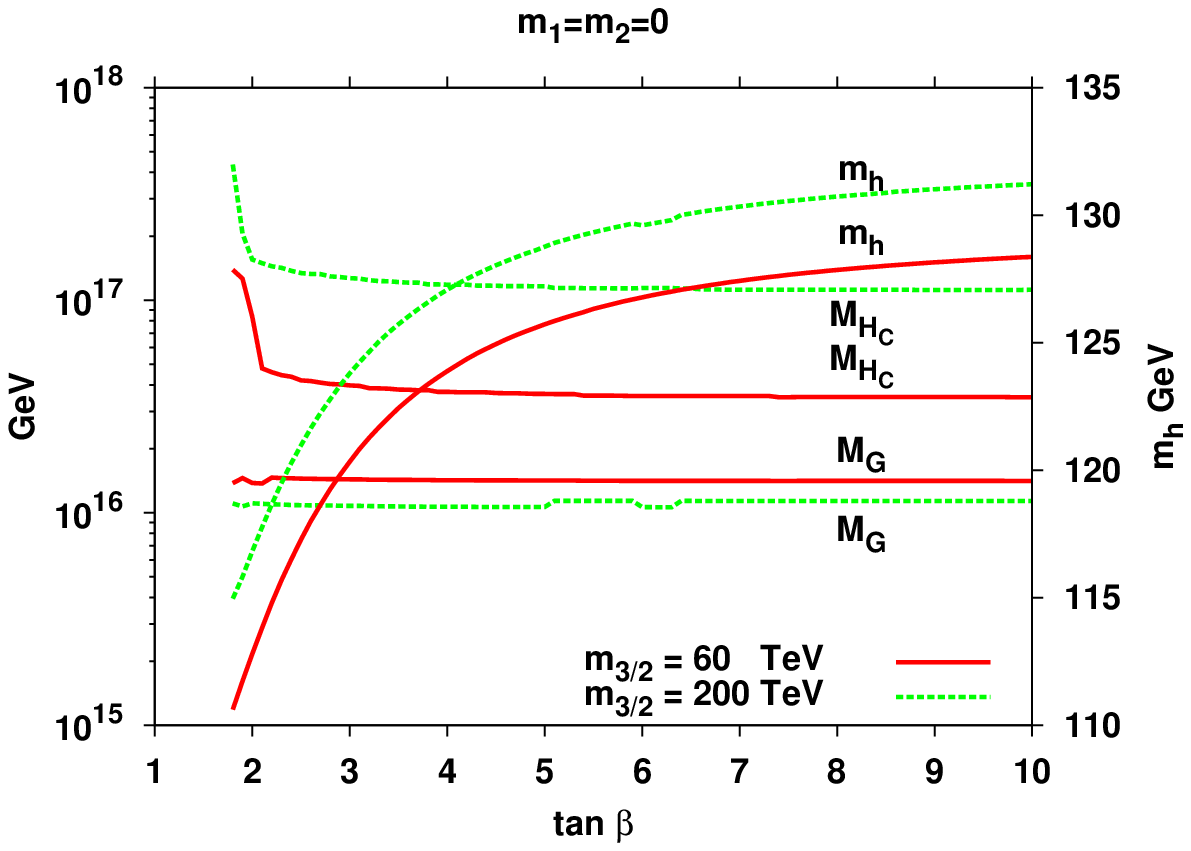}}
\scalebox{0.63}{\includegraphics{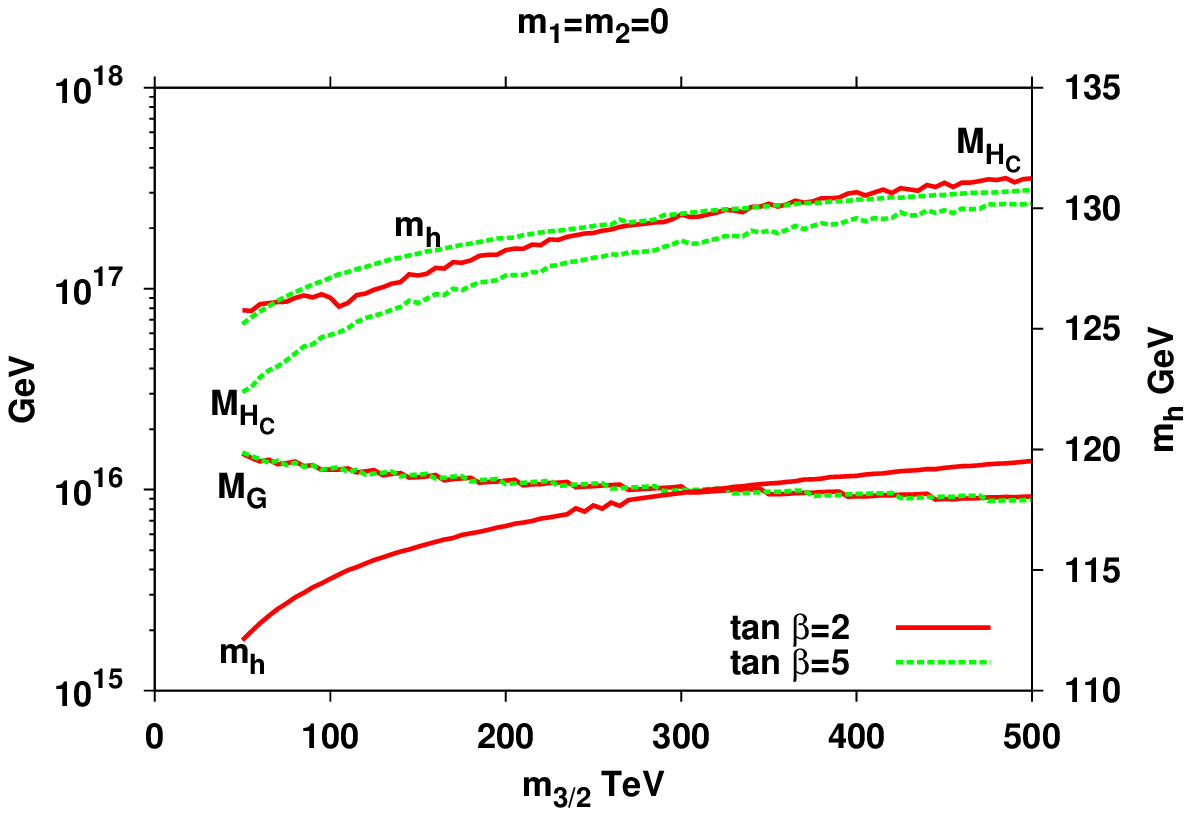}}
\caption{
{\it
The dependence of the GUT-scale mass parameters ($M_{H_C}$ and $M_G$) on
 $\tan \beta$ (left) for fixed $m_{3/2} = 60$ and $200$~TeV, and on
 $m_{3/2}$ (right) for fixed $\tan \beta = 2$ and $5$. Both of the Higgs
 soft masses have been fixed $m_1 = m_2 = 0$. The dependence of the
 Higgs mass is also shown with its value given on the right side of each
 panel. }}
\label{mm32tb}
\end{figure}

In contrast to the proton lifetime, the relevant GUT mass scales,
$M_{H_C}$ and $M_G$, are relatively insensitive to the PGM parameter choices
as seen in Fig.~\ref{mm32tb}. As one can see in the left panel, there is
very little dependence on $\tan \beta$.
The mass parameter $M_G\equiv (M_X^2M_\Sigma)^{\frac{1}{3}}$ is always
close to $10^{16}$~GeV independent of $m_{3/2}$ (as also seen in the
right panel).  While the color-triplet mass is insensitive to $\tan
\beta$, it does have a mild dependence on the gravitino mass
and ranges from a few $\times 10^{16}$ -- few $\times
10^{17}$~GeV. Notice that in the weak-scale SUSY scenario the mass of
the color-triplet Higgs multiplet is predicted to be around $10^{15}$~GeV
\cite{Murayama:2001ur}. A heavier color triplet mass makes the proton
lifetime long enough to evade the current experimental bound. Furthermore,
in some of the parameter space of PGM, the GUT-scale parameters
$M_{H_C}$ and $M_G$ are both of ${\cal O}(10^{16})$. In these cases, the threshold corrections at the GUT scale become very small,
which implies the unification of the gauge couplings is quite good. In fact, for $m_{3/2}\sim 60 $ TeV and $\tan\beta \sim 5$, we get good gauge coupling unification and a proton lifetime which could be in reach of future experiments.

\begin{figure}[t]
\scalebox{0.5}{\includegraphics{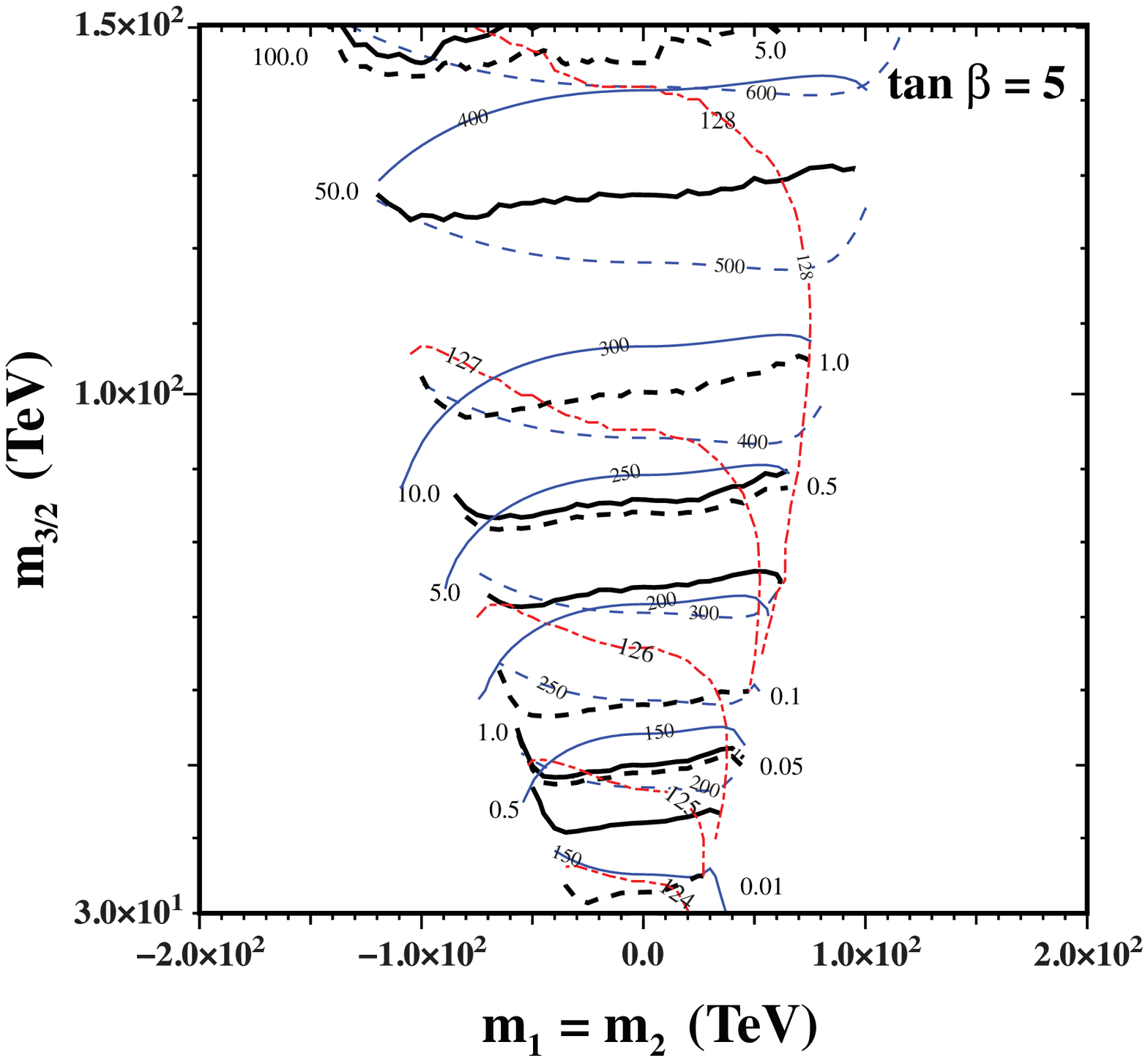}}
\scalebox{0.5}{\includegraphics{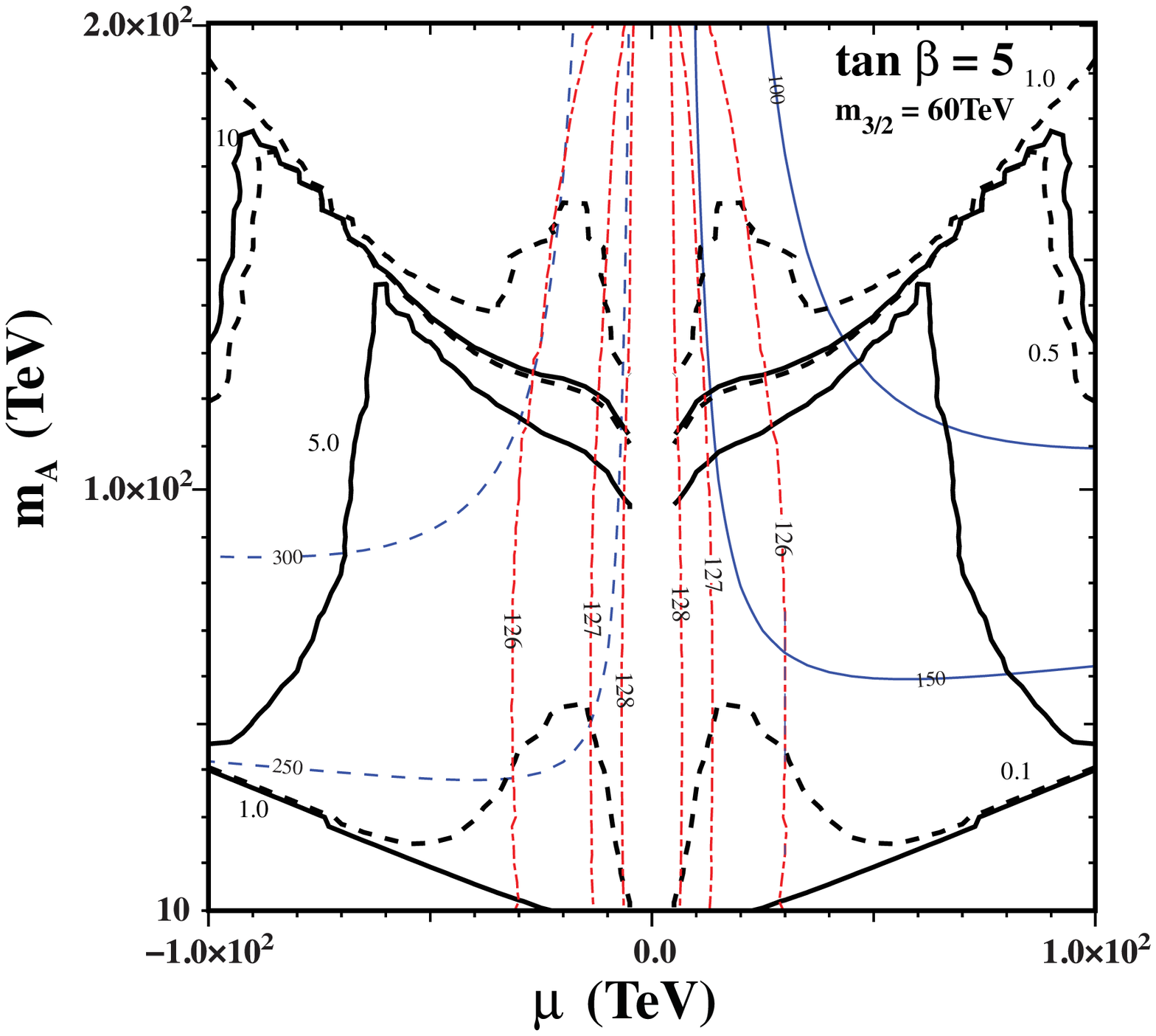}}
\caption{
{\it
a) The $m_{3/2}$--$m_{1,2}$ plane for $\tan \beta = 5$ and b)  the $\mu$--$m_A$ plane for
$\tan \beta = 5$ and $m_{3/2} = 60$ TeV. In both panels,
the Higgs mass is shown by thin red dot-dashed contours in $1$\,GeV intervals.
The wino/chargino mass is shown by the thin solid ($\mu > 0$) and dashed
($\mu < 0$) contours. The thick black contours show the value of the proton lifetime
based on the quark Yukawa couplings (solid) and lepton Yukawa couplings (dashed) in
units of $10^{35}$ years. Lifetime contours for the solid curves are labeled to the left of the contours
whereas dashed contours are labeled to the right.}}
\label{muma}
\end{figure}

In Fig.~\ref{muma}, we offer two additional planes which show the dependence of the proton lifetime
on other PGM parameters. In the left panel, we plot the lifetime contours in the $m_1=m_2$, $m_{3/2}$ plane.
This is again a NUHM1-like model and we have fixed $\tan \beta =5$. As in Fig.~\ref{m1eqm2p},
the red-dashed curves show the Higgs mass contours which vary from about 124--128 GeV for the plane shown.  As before, the curves extend across a limited range in $m_1=m_2$ where the EWSB conditions can be
satisfied. At large positive $m_1^2$, $\mu^2$ goes to 0 (where the curve is cutoff). At very small
$\mu$, the Higgs masses increases rapidly causing the sudden downturn in the mass contours.
As expected, we see the wino mass varies considerably as $m_{3/2}$ is varied.
For the range in $m_{3/2}$ shown, the proton lifetime varies from as low as $10^{33}$ years
using the lepton Yukawas and low $m_{3/2}$ to as high as $10^{37}$ years using quark Yukawas and
$m_{3/2} \approx 150$ TeV.

In the right panel of Fig.~\ref{muma}, we show a two-parameter extension of the two-parameter
PGM similar to the NUHM2 \cite{nuhm2}. Results are displayed in the $\mu,m_A$ plane for fixed
$\tan \beta = 5$ and $m_{3/2} = 60$ TeV.
In this case, the EWSB conditions, are used to solve for the two Higgs soft masses which now differ.
As the Higgs mass is largely independent of $m_A$, the Higgs mass contours are nearly vertical.
At the center of the plot, as $|\mu|$ gets to be very small, $m_h$ gets large and exceeds 130 GeV.
At large $|\mu|$, $m_h$ is always larger than 125 GeV in the ranges shown. The threshold
corrections to the wino mass are sensitive to $\mu$ and $m_A$
and that accounts for the variation of $m_\chi$ as these parameters are varied.

\begin{figure}[t!]
\scalebox{0.63}{\includegraphics{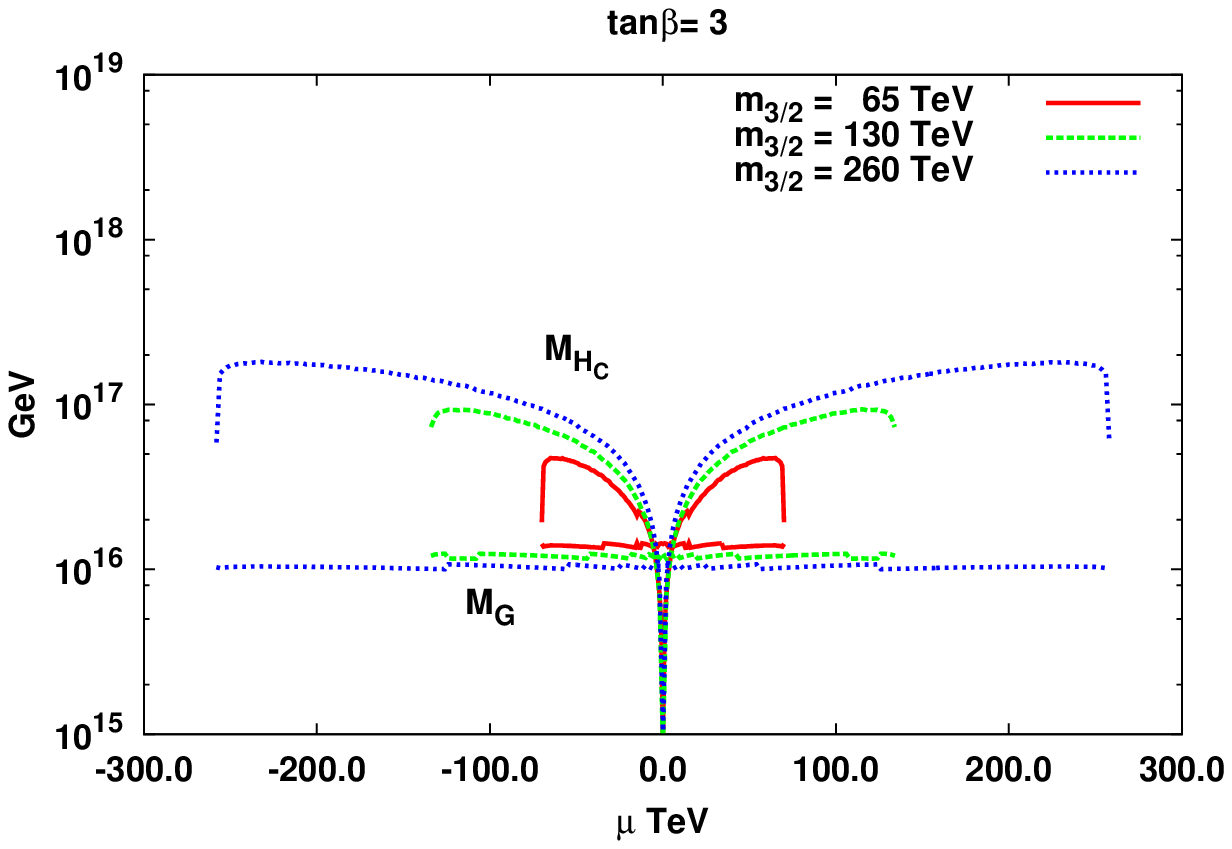}}
\scalebox{0.63}{\includegraphics{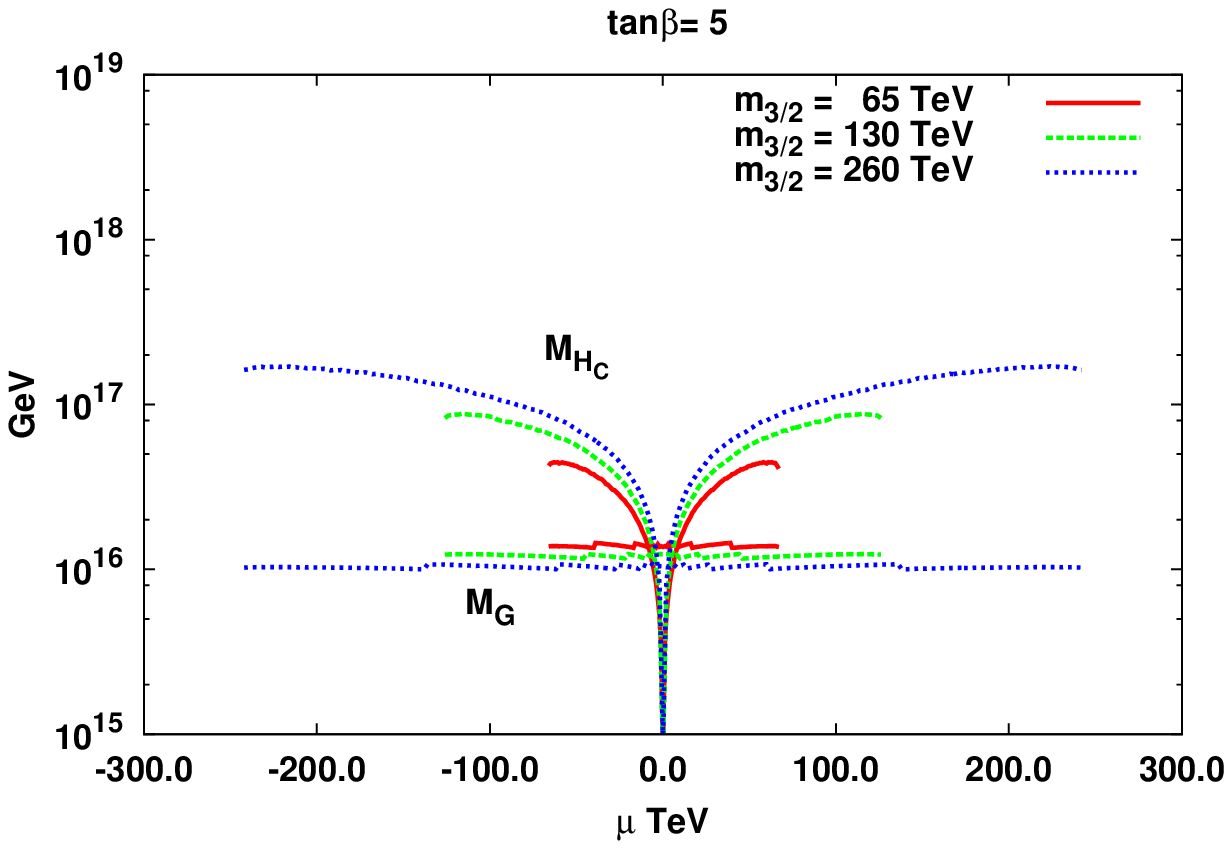}}
\caption{
{\it
The dependence of the GUT-scale masses ($m_{H_C}$ and $M_G$) on $\mu$ for
fixed $\tan \beta = 3$  (left) and $\tan \beta = 5$ (right) and fixed $m_{3/2} = 65, 130$, and $260$ TeV.
Both of the Higgs soft masses have been fixed $m_1 = m_2 = 0$. The extent of the curves is
determined by the validity of radiative EWSB.
}}
\label{tpmu}
\end{figure}

\begin{figure}[t!]
\scalebox{0.63}{\includegraphics{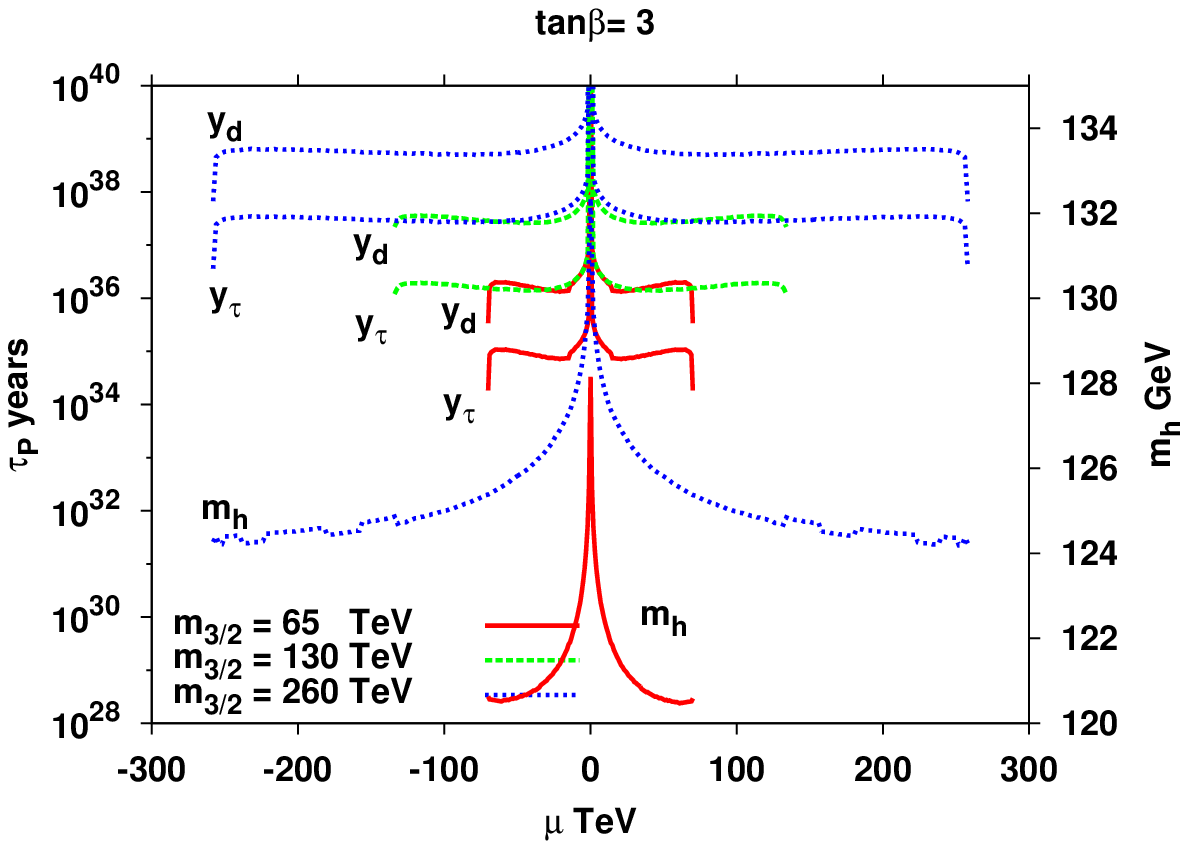}}
\scalebox{0.63}{\includegraphics{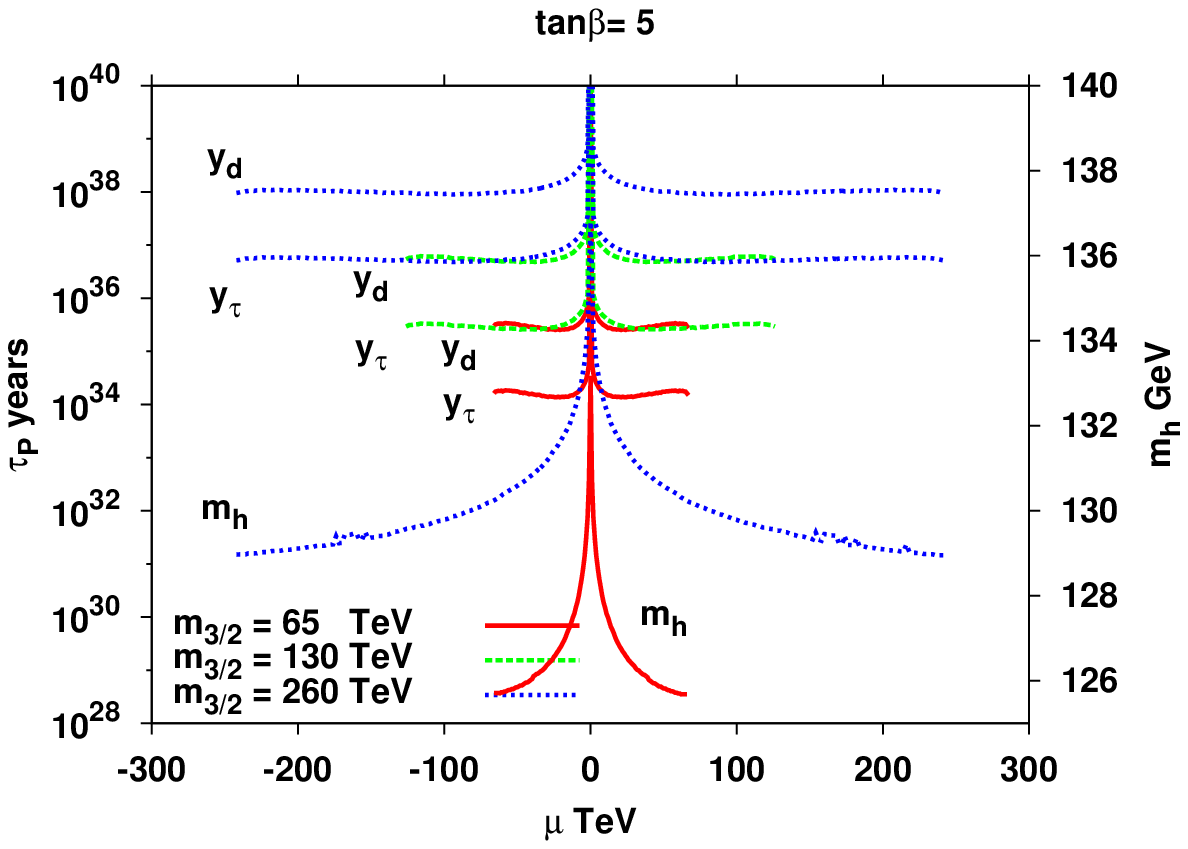}}
\caption{
{\it
The dependence of the proton lifetime on $\mu$ for
fixed $\tan \beta = 3$  (left) and $\tan \beta = 5$ (right) and fixed $m_{3/2} = 65, 130$, and $260$~TeV.
Both of the Higgs soft masses have been fixed $m_1 = m_2 = 0$. The extent of the curves is
determined by the validity of radiative EWSB.
The dependence of the Higgs mass is also shown with its value given on the right side of each panel.  }}
\label{mmu}
\end{figure}

The proton lifetime varies between $10^{34}$ and $10^{36}$ years but
shows significantly more variability.  This is due to the competing effects
of changing $\mu$. The proton lifetime depends both on the color-triplet
Higgs mass and on $\mu$ itself.\footnote{The proton decay rate directly
depends on $\mu$ through the loop function $F$ in
Eq.~\eqref{eq:susymatching}. When $|\mu| << m_{3/2}$, $F\propto \mu
/m_{3/2}^2$, while if $\mu >> m_{3/2}$, $F\propto
\log(\mu^2/m_{3/2}^2)/\mu$, as can be seen from Eq.~\eqref{eq:funceq}.  } As $\mu$ is lowered, the color-triplet Higgs
mass decreases which tends to decrease the proton lifetime.  But as $|\mu|$ is further decreased,
the proton lifetime dependence on $\mu$ overcomes its dependence on $M_{H_C}$ and the lifetime
increases very rapidly at small $|\mu|$ seen by the sharp downturn in the contours near $\mu = 0$.
These effects can be better understood by examining Figs.~\ref{tpmu} and \ref{mmu} which show
the behavior of the proton lifetime and GUT-scale masses, including the heavy Higgs mass, as a function of $\mu$ for fixed $\tan \beta$ and
$m_{3/2}$. Here we see the first gradual and then rapid decrease in the color-triplet mass as
$|\mu|$ is lowered from large values toward $\mu = 0$. There is no substantial difference in this
behavior between the two values of $\tan \beta$ shown.
Once again, we see that $M_G$ depends very little on our parameter
choices and is always near $10^{16}$~GeV.

Finally, in Fig.~\ref{mmu}, we see the sharp increase in the proton lifetime
as $|\mu|$ gets small.\footnote{Our calculations are only valid for $|\mu|$ much greater than the wino mass.} Here we see also that the Higgs mass rises sharply as $\mu$ tends to
zero.  It is important to recall that the lifetime plotted corresponds only to that
given by the dimension-five operator given in Eq.~\eqref{tp5} and would not exceed
$3 \times 10^{35}$ years when the dimension-six operator is included. The latter is fairly insensitive
to parameter choices.

\section{Conclusion and Discussion}

As we await new results for physics beyond the standard model from the LHC,
we have been forced to consider supersymmetric models with sfermion masses larger than
what was previously considered `natural'. While a great deal of attention had been focused
on relatively simple models such as the CMSSM or mSUGRA (with four and three parameters respectively)
or the NUHM1,2 with five and six parameters, pure gravity mediation models can be
described with as few as two parameters at the cost of a mass spectrum which approaches
the PeV scale. As we hope the actual theory of nature is in the realm of experimental science,
it is imperative to find means to test these models. Here we have examined
one additional possibility for testing these models despite their
generally heavy mass spectra.

PGM theories, with all their economy, are still able to resolve many of the questions
their lower energy cousins (such as the CMSSM) were motivated from. These include
the ability to achieve gauge coupling unification at the GUT scale, radiative breaking
of the electroweak symmetry, the stability of the Higgs potential, and they also provide a suitable
candidate for dark matter.  The latter is definitely more difficult in PGM models, as
the wino is usually the lightest supersymmetric particle and as such would require
a wino mass near 3 TeV to supply the correct relic density. This pushes the gravitino mass
up to several hundred TeV. Alternatives within PGM are possible if $\mu$ is relatively small
and the Higgsino is the lightest supersymmetric particle \cite{eioy5} or if the theory contains
additional vector-like states and bino-gluino co-annihilation controls the relic bino density \cite{vector},
or even axion dark matter \cite{eioy5,eioy4}. In contrast to their lower energy counterparts,
PGM models have a relatively easy time obtaining a Higgs mass in agreement with the
experimental measurement \cite{lhch}.

Thus experimental verification of PGM models remains challenging.
While there is the chance that the lightest supersymmetric particle is within
reach of the LHC, the bulk of the PGM spectrum is not. Here we have calculated
the proton lifetime in PGM models. We have found that typically
the lifetime is long and in many cases significantly above the current
experimental bounds. However in cases where $m_{3/2}$ is relatively small
and $\tan \beta$ is relatively high, the proton lifetime is low and
may be at the level of current experimental searches. While proton decay itself,
can not point directly to PGM supersymmetry, it may provide one more handle on
an ever increasingly elusive theory beyond the standard model.

\section*{Acknowledgments}

The work of J.E. and K.A.O. was supported in part
by DOE grant DE-SC0011842 at the University of Minnesota.
The work of N.N. is supported by Research Fellowships of the Japan Society
for the Promotion of Science for Young Scientists.

\section*{Appendix}
\appendix
\section{Minimal SU(5) Notation and Conventions
\label{App:NotCon}}

Here, we review the minimal SUSY SU(5) GUT
\cite{Dimopoulos:1981zb, Sakai:1981gr} and clarify our notation and
conventions. In these models, the MSSM matter fields are embedded into
a  $\bar{\bf 5}$ and ${\bf 10}$ representations of the SU(5)
gauge group for each generation. Let $\Phi_i$ and $\Psi_i$ be the chiral
superfields in the $\bar{\bf 5}$ and ${\bf 10}$ representations,
respectively, with $i$ denoting the generation index. These fields decompose into the
MSSM superfields as
\begin{align}
 \Phi_i &=
\begin{pmatrix}
 \bar{D}_{i1} \\
 \bar{D}_{i2} \\
 \bar{D}_{i3} \\
 E_i \\
 -N_i
\end{pmatrix}~, ~~~~~~
\Psi_i=\frac{1}{\sqrt{2}}
\begin{pmatrix}
 0&\bar{U}_{i3}&-\bar{U}_{i2}&U^{1}_i&D^{1}_i \\
 -\bar{U}_{i3}&0&\bar{U}_{i1}&U^{2}_i&D^{2}_i\\
 \bar{U}_{i2}&-\bar{U}_{i1}&0&U^{3}_i&D^{3}_i\\
 -U^{1}_i&-U^{2}_i&-U^{3}_i&0&\bar{E}_i \\
 -D^{1}_i&-D^{2}_i&-D^{3}_i&-\bar{E}_i &0
\end{pmatrix}~,
\end{align}
with
\begin{equation}
 L_i=
\begin{pmatrix}
 N_i \\ E_i
\end{pmatrix}~, ~~~~~~
Q^a_i=
\begin{pmatrix}
 U^a_i \\ D^a_i
\end{pmatrix}~,
\end{equation}
where $a=1,2,3$ denotes the color index.
The MSSM Higgs superfields, on the other hand, are embedded
into a {\bf 5} and $\bar{\bf 5}$:
\begin{equation}
 H=
\begin{pmatrix}
 H^1_C \\ H^2_C \\ H^3_C \\ H^+_2 \\ H^0_2
\end{pmatrix}
,~~~~~~\bar{H}=
\begin{pmatrix}
 \bar{H}_{C1}\\
 \bar{H}_{C2}\\
 \bar{H}_{C3}\\
 H^-_1 \\ -H^0_1
\end{pmatrix}
~,
\end{equation}
where the last two components are the MSSM Higgs superfields,
\begin{equation}
 H_2=
\begin{pmatrix}
 H^+_2 \\ H^0_2
\end{pmatrix}
,~~~~~~H_1=
\begin{pmatrix}
 H^0_1 \\ H^-_1
\end{pmatrix}
~.
\end{equation}
The other piece of the ${\bf 5}$ and $\bar{\bf 5}$ Higgs bosons are labeled by
$H^a_C$ and $\bar{H}_{Ca}$ and will be referred to as the color-triplet
Higgs bosons.

The gauge boson of SU(5) is a ${\bf 24}$. In supersymmetry this
corresponds to a real vector superfield, ${\cal V}^A$, where $A=1,\dots,
24$ represents the gauge index. $V^A$ can be decomposed into the SM
gauge fields, plus the additional massive gauge bosons of SU(5)
breaking, as follows
\begin{equation}
 {\cal V}\equiv {\cal V}^AT^A=\frac{1}{\sqrt{2}}
\begin{pmatrix}
 \begin{matrix}
G  -\frac{2}{\sqrt{30}} B
 \end{matrix}
&
\begin{matrix}
X^{\dagger 1} \\
X^{\dagger 2} \\
X^{\dagger 3}
\end{matrix}
&
\begin{matrix}
 Y^{\dagger 1}\\ Y^{\dagger 2} \\ Y^{\dagger 3}
\end{matrix}
\\
\begin{matrix}
 X_1 & X_2 & X_3 \\
 Y_1 & Y_2 & Y_3
\end{matrix}
&
\begin{matrix}
 \frac{1}{\sqrt{2}}W^3+\frac{3}{\sqrt{30}}B \\ W^-
\end{matrix}
&
\begin{matrix}
W^+ \\ - \frac{1}{\sqrt{2}}W^3+\frac{3}{\sqrt{30}}B
\end{matrix}
\end{pmatrix}
~,
\end{equation}
where $T^A$ is the generator of the fundamental representation of the SU(5),
and $G$, $B$, and $W$ denote the MSSM gauge vector superfields with
there associated generators. The massive gauge bosons associated with
the breaking of SU(5) typically referred to as $X_a$ and $Y_a$ will be
called just the $X$-bosons with definition
\begin{equation}
 (X)^\alpha_a =
\begin{pmatrix}
 X^1_a \\ X^2_a
\end{pmatrix}
\equiv
\begin{pmatrix}
 X_a \\ Y_a
\end{pmatrix}~.
\end{equation}
Here $\alpha,\beta,\dots$ denote the SU(2)$_L$ indices.

The simplest means of breaking SU(5) to the SM gauge symmetries
$\text{SU}(3)_C \otimes \text{SU}(2)_{L} \otimes \text{U}(1)_Y$ is via
an adjoint ${\bf 24}$ discussed in the text.  The ${\bf 24}$ decomposes
as follows:
\begin{equation}
\Sigma \equiv
\Sigma^A T^A
\begin{pmatrix}
 \Sigma_8&\Sigma_{(3,2)} \\
 \Sigma_{(3^*,2)} & \Sigma_3
\end{pmatrix}
+\frac{1}{2\sqrt{15}}
\begin{pmatrix}
 2&0\\0&-3
\end{pmatrix}
\Sigma_{24}~.
\end{equation}
Without losing any generality and for simplicity, we assume all SU(5)
breaking occurs along the $\Sigma_{24}$ direction which is separated in
the above equation.

\section{Proton Decay \label{App:ProDec}}

In this appendix, we give additional details of our calculation of the
proton lifetime. The important Wilson coefficients arising from
integrating out the colored Higgs triplet are
\begin{align}
 C^{3311}_{5R}
(Q_G)=\frac{1}{M_{H_C}}f_tf_d(Q_G)
e^{-i\varphi_1}V_{tb}V_{ud}^*~, \nonumber \\
 C^{3312}_{5R}(Q_G)=\frac{1}{M_{H_C}}f_tf_s(Q_G)
e^{-i\varphi_1}V_{tb}V_{us}^*~.\label{eq:wilson3gen}
\end{align}
These coefficients are then evolved down to the SUSY scale using
\begin{equation}
 \frac{d}{d\ln Q} C_{5R}^{331l} =
\frac{1}{16\pi^2}\biggl[
-\frac{12}{5}g_1^2-8g_3^2 +2f_t^2 +2f_\tau^2
\biggr]C_{5R}^{331l}~,
\end{equation}
where $l=1,2$ and $Q$ is the renormalization scale.

\begin{figure}[t]
\begin{center}
\includegraphics[height=60mm]{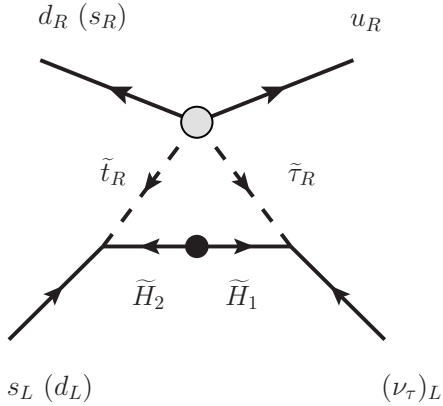}
\caption{{\it One-loop Higgsino-exchanging diagram which gives rise to the
 dominant contribution to the baryon-number violating
 four-Fermi operators. Gray dot indicates the
 dimension-five effective interaction, while black dot represents
 the  Higgsino mass term.}}
\label{fig:1loop}
\end{center}
\end{figure}

At the SUSY scale
$Q_S$, the sfermions are integrated out via the diagram in
Fig.~\ref{fig:1loop} to give
\begin{equation}
 {\cal L}^{\text{eff}}_6=C_i~ \epsilon_{abc}(u^a_{R1}d^b_{Ri})
(Q_{L3}^c \cdot L_{L3})~,
\end{equation}
with
\begin{equation}
 C_i (Q_S)=\frac{f_tf_\tau}{(4\pi)^2}C^{*331i}_{5R}(Q_S)
F(\mu, m_{\widetilde{t}_R}^2,m_{\tau_R}^2)~,
\end{equation}
where $i=1,2$ and
 \begin{align}
F(M, m_1^2, m_2^2) &\equiv
\frac{M}{m_1^2-m_2^2}
\biggl[
\frac{m_1^2}{m_1^2-M^2}\ln \biggl(\frac{m_1^2}{M^2}\biggr)
-\frac{m_2^2}{m_2^2-M^2}\ln \biggl(\frac{m_2^2}{M^2}\biggr)
\biggr]~.
\label{eq:funceq}
\end{align}

These Wilson coefficients $C_i$, which are initially defined at the SUSY
scale, are then run down from the weak scale using \cite{Abbott:1980zj}
\begin{equation}
 \frac{d}{d\ln Q}C_i =
\biggl[\frac{\alpha_1}{4\pi}\biggl(-\frac{11}{10}\biggr)
+\frac{\alpha_2}{4\pi}\biggl(-\frac{9}{2}\biggr)
+\frac{\alpha_3}{4\pi}(-4)
\biggr]C_i ~.
\end{equation}
At the weak scale the Lagrangian takes the form
\begin{equation}
 {\cal L}(p\to K^+\bar{\nu}_\tau)
=C_{usd} [\epsilon_{abc}(u_R^as_R^b)(d_L^c\nu_\tau)]
+C_{uds} [\epsilon_{abc}(u_R^ad_R^b)(s_L^c\nu_\tau)]
~,
\end{equation}
with
\begin{align}
 C_{usd}&= -V_{td}C_2(m_Z) ~, \nonumber \\
 C_{uds}&= -V_{ts}C_1(m_Z)~.
\end{align}
The new Wilson coefficients $C_{usd,uds}$ are then further run down to
the hadronic scale $Q_{\text{had}}=2$ GeV.  Below the electroweak scale,
the RGEs of the Wilson coefficients are
given by
\begin{equation}
 \frac{d}{d\ln Q}C_{usd,uds}
=-\biggl[
4 \frac{\alpha_s}{4\pi} +\biggl(\frac{4}{3}+\frac{4}{9}N_f\biggr)
\frac{\alpha_s^2}{(4\pi)^2}
\biggr]C_{usd,uds}~,
\end{equation}
at the two-loop level \cite{Nihei:1994tx}. The solution for this equation is
\begin{align}
 A_L
\equiv \frac{C(Q_{\text{had}})}{C(m_Z)}
=\biggl[
\frac{\alpha_s(Q_{\text{had}})}{\alpha_s(m_b)}
\biggr]^{\frac{6}{25}}\biggl[
\frac{\alpha_s(m_b)}{\alpha_s(m_Z)}
\biggr]^{\frac{6}{23}}
\biggl[
\frac{\alpha_s(Q_{\text{had}})+\frac{50\pi}{77}}
{\alpha_s(m_b)+\frac{50\pi}{77}}
\biggr]^{-\frac{173}{825}}
\biggl[
\frac{\alpha_s(m_b)+\frac{23\pi}{29}}
{\alpha_s(m_Z)+\frac{23\pi}{29}}
\biggr]^{-\frac{430}{2001}}~.
\end{align}
This long-range renormalization factor is computed to be $A_L = 1.247$ and appears as a multiplicative factor to the Wilson coefficients defined at the weak scale. The Wilson coefficients at the hadronic scale are then
\begin{eqnarray}
C_{usd,uds}(Q_{\text{had}})= C_{usd,uds}(m_Z) A_L ~.
\end{eqnarray}

The partial decay width for $p\to K^+ \bar{\nu}$ is then found to be
\begin{equation}
 \Gamma(p\to K^+\bar{\nu})=\frac{m_p}{32\pi}
\biggl(1-\frac{m_K^2}{m_p^2}\biggr)^2
|{\cal A}(p\to K^+\bar{\nu})|^2~,
\end{equation}
where $m_p$ and $m_K$ are the proton and kaon masses, respectively,
and
\begin{equation}
 {\cal A}(p\to K^+\bar{\nu})=
C_{usd}(Q_{\text{had}})\langle K^+\vert (us)_R ^{}d_L^{}\vert p\rangle +
C_{uds}(Q_{\text{had}})\langle K^+\vert (ud)_R ^{}s_L^{}\vert p\rangle ~.
\end{equation}

{}


\end{document}